\documentclass[aps,prl,twocolumn, superscriptaddress]{revtex4-1}

\usepackage{graphicx}
\usepackage{dcolumn}
\usepackage{bm}

\begin{document}

\preprint{APS/123-QED}

\title{A Turing model of pattern formation in atmospheric pressure gas discharges}

\author{Xi Chen}
\altaffiliation[Also at ]{the Department of Physics, Tsinghua University.}
\author{Yao Zhou}
\altaffiliation[Also at ]{the Department of Physics, Tsinghua University.}
\author{Xi-Ming Zhu}
\author{Yi-Kang Pu}
\affiliation{Department of Engineering Physics, Tsinghua University, Beijing 100084, China}
\author{F. Iza}
\affiliation{Department of Electronic and Electrical Engineering, Loughborough University, Leicestershire, LE11 3TU, UK}
\author{M. A. Lieberman}
\affiliation{Department of Electrical Engineering and Computer Science, University of California, Berkeley, California 94720, USA}

\date{\today}

\begin{abstract}
In this letter we propose a Turing model of the formation of patterns of visible light emission intensity in atmospheric pressure gas discharges. The electron density and the electron temperature take the roles of activator and inhibitor respectively in a two-reactant Turing model, with the activator diffusion coefficient being much smaller than that of the inhibitor, and ionization and excitation from excited state atoms considered as the dominant reaction processes. The model predicts striations in a 1D system, which quantitatively agree with experimental results in terms of the spatial variation scale lengths. Additionally, the model also predicts changes in the discharge structure observed experimentally when input power and gas pressure are varied.
\end{abstract}


\maketitle

Plasmas, especially those at atmospheric pressure, often exhibit non-uniform and/or non-stationary emission intensity of visible light. These phenomena, called patterns, are common in non-equilibrium systems \cite{Nicolis,Cross,Cross2}. Patterns including striations, constrictions, zigzags, hexagons, clusters, etc. have been observed in gas discharges since the 1830s \cite{Massey,Kabouzi,Kolobov,Ammelt,Astrov,Astrov2,Ammelt3}. Models based on skin effects \cite{Massey,Kabouzi}, non-uniform gas heating \cite{Rogoff,Kabouzi2} and electron-electron collisions \cite{Golubovskii} were proposed to explain these phenomena, but all these interpretations could not explain some of the important characteristics of the patterns such as the discharge instabilities and the spatial periodicity. Simulations based on the kinetic theory successfully reproduced striations \cite{Kolobov} and constrictions \cite{Dyatko}, but the basic mechanism of the pattern formation is not revealed due to the many complicated processes included in the simulations. 

In 1952, A. M. Turing adopted bifurcation analysis of coupled nonlinear reaction-diffusion equations to explain pattern formation in reaction-diffusion systems \cite{Turing}, which is now referred to as the Turing model. A typical two-reactant Turing model includes an activator-inhibitor system, where the diffusion coefficient of the inhibitor should be much larger than that of the activator \cite{Cross2}. The model was later applied to explain patterns in DC discharges \cite{Radehaus,Radehaus2,Ammelt,Ammelt3,Astrov,Astrov2}, with current density and electric potential chosen as reactants. However, the theory is limited to the two-layer discharge structure of the system, and thus could not explain the patterns reported recently in other gas discharge systems, including microwave plasmas \cite{Iza}, dielectric barrier discharges \cite{Guikema}, boundary layer discharges \cite{Schoenbach} and breakdown processes \cite{Hidaka}. These patterns share similar characteristics, for instance, when the pressure is decreased or the power is raised, the discharge becomes uniform \cite{Iza,Guikema,Schoenbach}. Such consistency suggests that there should be a common physical mechanism for these patterns. 

In this letter we introduce the electron density and the electron temperature as the activator and the inhibitor respectively into a Turing model to explain pattern formation in gas discharges. These two reactants are fundamental parameters controlling the various processes in plasmas. While this model can be applied to various discharge systems, we use a particular one at atmospheric pressure as an example here to illustrate that the predictions of the model can indeed be quantitatively consistent with experimental results. 

The experimental observations considered here are the striations of the visible light emission reported in atmospheric pressure argon discharges generated with a microwave induced plasma source based on a microstrip split-ring resonator \cite{Iza}, where the striations are distributed periodically along the gap between the electrodes. Some experimental images are shown in FIG. \ref{fig3}, (a)-(c). This periodic distribution of the visible light emission intensity suggests a periodic distribution of the excited state atom density, which depends on the electron density $n$ and the electron temperature $T$. So it is implied that their distributions should also be non-uniform. To describe such non-uniformity, we first consider the electron continuity equation:
\begin{equation}
\frac{\partial n}{\partial t}=D_a\frac{\partial^2n}{\partial x^2} +G_n-L_n\label{model1} 
\end{equation}
Here $x$ denotes the position along the gap, and $D_a$ is the ambipolar diffusion coefficient, while $G_n$ and $L_n$ are the production and loss rates of the electron density respectively. At atmospheric pressure, the ionization from the excited state atoms dominate the production of the electron density, and the collisional-radiative model suggests that the production rate is \cite{Zhu}:
\begin{equation}\label{gainn}
G_n=r_1n^2e^{-E_1/T}
\end{equation}
Here $r_1$ is a constant and $E_1$ is an effective ionization energy for excited state atoms.
In atmospheric pressure argon discharges, the electron loss is mainly due to electron recombination with argon dimer ions: $Ar^++Ar+Ar\to Ar+Ar_2^+$, $e+Ar_{2}^{+} \to Ar+Ar$ \cite{Zhu}. The rates of these reactions can be expressed as $R_1=r_2n(Ar^+)$ and $R_2=L_n$ respectively, with $r_2$ being a constant. These two processes are also the dominant production and loss processes of $Ar_2^+$, which are locally balanced: $R_1=R_2$. In addition, according to the collisional-radiative model \cite{Zhu}, the density of $Ar^+$ is much higher than that of $Ar_2^+$, thus approximately equal to the electron density: $n(Ar^+)=n$. Then we have:
\begin{equation}\label{lossn}
L_n=R_2=R_1=r_2n(Ar^+)=r_2n
\end{equation}

The patterns in gas discharges often come with a non-uniform distribution of the electron temperature \cite{Kolobov}, the existence of which has been proved possible in the discharge system discussed here \cite{Iza2,Zhu2}. To describe such a no-uniformity, we start with the electron energy conservation equation:
\begin{equation}
\frac{\partial}{\partial t}\left(\frac{3}{2}p\right)+\frac{\partial}{\partial x}\left(\frac{3}{2}pu\right)+p\frac{\partial u}{\partial x}-\frac{\partial}{\partial x}(\kappa_T\frac{\partial T}{\partial x})=G_E-L_E \label{energy}
\end{equation}
Here $p$ is the pressure, and we assume the energy distribution of the electrons to be Maxwellian: $p=nT$. $u$ is the flow velocity, and $\kappa_T=(3/2)nD_e$ is the thermal conductivity, where $D_e$ is the diffusion coefficient of the electrons. $G_E$ and $L_E$ denote the gain and loss rate of the electron energy respectively.
In the ambipolar regime, $D_e$ is $2\sim3$ orders of magnitude larger than $D_a$. So, compared with the heat conduction term, the energy flux caused by the flow, that is, terms containing $u$, can be neglected. 
In the bulk plasma, $G_E$ is mainly caused by the Ohmic heating \cite{Lieberman}. When filaments form, more power is coupled to these channels of high conductivity (high electron density) than to the neighboring regions of lower conductivity (lower eletron density). In a first approximation, this non-uniform power deposition can be modeled by assuming that the power density is proportional to the plasma conductivity, and thereby, the electron density:
\begin{equation}\label{gainE}
G_E=(3/2)r_3n
\end{equation}
Here $r_3$ is a constant that depends on the input power. 
As to $L_E$, the excitation of excited state atoms dominates the loss of the electron energy \cite{Zhu}:
\begin{equation}\label{lossE}
L_E=(3/2)r_4n^2e^{-E_2/T}
\end{equation}
Here $r_4$ is another constant, and $E_2$ is an effective excitation energy, with $E_2<E_1$. Based on the considerations above, (\ref{energy}) becomes:
\begin{equation}
T\frac{\partial n}{\partial t}+n\frac{\partial T}{\partial t}-D_e\frac{\partial n}{\partial x}\frac{\partial T}{\partial x}-nD_e\frac{\partial^2T}{\partial x^2}=r_3n-r_4n^2e^{-E_2/T}\label{diffusionTprime}
\end{equation}
The time derivative of the electron density in the first term of the LHS is given by (\ref{model1}) and can be neglected because $D_e\gg D_a$, $r_3\gg Tr_2$, and $r_4e^{-E_2/T}\gg Tr_1e^{-E_1/T}$ \cite{Zhu}. We also neglect the third term on the LHS because a bifurcation analysis \cite{Turing,Cross2} shows that this term does not affect the periodicity of the striations or the phase diagram of discharge transitions (which will be discussed later). Neglecting the two terms in (\ref{diffusionTprime}), dividing the equation by $n$, and substituting (\ref{gainn}) and (\ref{lossn}) into equation (\ref{model1}), we obtain a set of coupled nonlinear reaction-diffusion equations for the electron density and the electron temperature:
\begin{eqnarray}
&&\frac{\partial n}{\partial t}=D_a\frac{\partial^2n}{\partial x^2} +r_1n^2e^{-E_1/T}-r_2n\label{modeln} \\
&&\frac{\partial T}{\partial t}=D_e\frac{\partial^2T}{\partial x^2}+r_3-r_4ne^{-E_2/T}  \label{modelT}
\end{eqnarray}

\begin{figure}
 \includegraphics[scale=0.4]{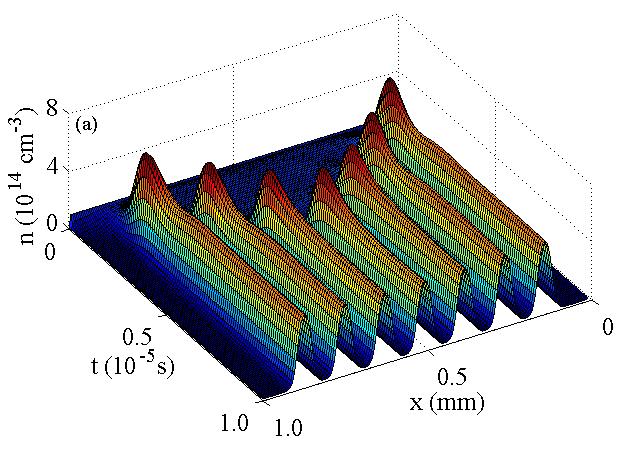}
  \includegraphics[scale=0.4]{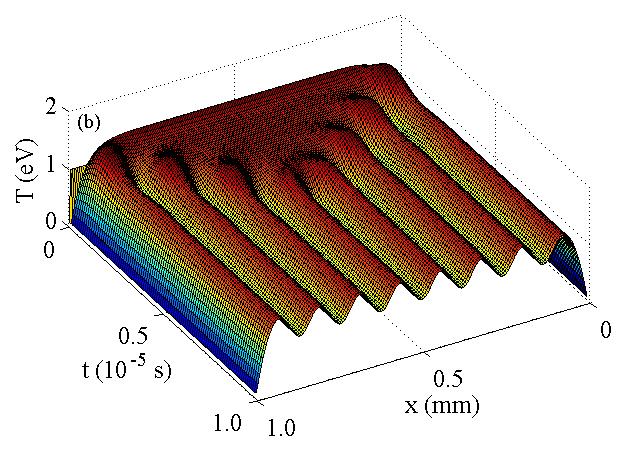}
 \caption{\label{fig1}Spatio-temporal evolutions of the electron density (a) and the electron temperature (b) distributions. The discharge condition is 1 atm, 1 W.}
 \end{figure}
 
In equations (\ref{modeln}) and (\ref{modelT}), which describe the gas discharge system, the electron density promotes its own production via stepwise ionization, while the electron temperature self-inhibits: a perturbation that increases the electron temperature results in increased energy losses due to the increase in excitation processes (third term on the RHS of equation (\ref{modelT})), driving the electron temperature down and ``inhibiting" the initial perturbation. In addition, $D_e\gg D_a$. So, the two reactants satisfy the conditions of a typical Turing activator-inhibitor system, and thus patterns can result from the dynamics of the system.

With the parameters of an atmospheric pressure argon discharge \cite{Zhu}, including reaction rates, ionization and excitation threshold energies, the equations (\ref{modeln}) and (\ref{modelT}) are solved numerically under uniform initial conditions ($10^{14}$ cm$^{-3}$, 1 eV) and fixed boundary conditions ($10^{13}$ cm$^{-3}$, 0.1 eV) (in fact the boundary and initial conditions have little influence on major quantitative characteristics of the final striations, including periods and amplitudes). The spatio-temporal  evolutions of the electron density and the electron temperature obtained are shown in FIG. \ref{fig1}. The profiles evolve from uniform at the beginning to steady periodic structures at the end, which can result in striations of visible light emission intensity shown in FIG. \ref{fig3}, (B). The opposite phase distributions of the electron density and the electron temperature shown in FIG. \ref{fig1} are consistent with simulation results based on the kinetic theory \cite{Kolobov}. The magnitude of the spatial period of the striations, which has an order of $\sim0.1$ mm, agrees with the experimental results (FIG. \ref{fig3}, (b)).

It has been observed in experiments that the number of striations increases with the input power (FIG. \ref{fig3}, (b) and (c)). This behavior is also predicted by the model: with bifurcation analysis \cite{Cross,Cross2}, the spatial period of the striations $\lambda$ can be determined with the parameters in (\ref{modeln}) and (\ref{modelT}):
\begin{equation}
\lambda=2\pi\left\{\frac{r_2r_3}{D_eD_a}\cdot\frac{[\log(r_1r_3 /r_2r_4)]^2}{E_1-E_2}\right\}^{-1/4} \label{period}
\end{equation}
Note that the spatial period of the striations decreases when $r_3$, which represents the Ohmic heating term (\ref{gainE}), increases. In other words, the higher the input power, the more striations exist.

In experiments, it has also been observed that when the pressure decreases, the discharge structure changes from periodic to uniform, i.e. striations disappear (FIG. \ref{fig3}, (a)). Such phenomena can also be explained with the model:
bifurcation analysis \cite{Cross,Cross2} also suggests that the formation of the striations demands the following two conditions:
\begin{eqnarray}
\frac{D_e}{D_a}r_2-\frac{E_2}{T_0^2}r_3&>&2\sqrt{\frac{D_e}{D_a}\cdot\frac{E_1-E_2}{T_0^2}\cdot r_2r_3} \label{inequality2}\\
r_2-\frac{E_2}{T_0^2}r_3&<&0 \label{inequality1}
\end{eqnarray}
Here $T_0=(E_1-E_2)/[\log(r_1r_3 /r_2r_4)]$.
When (\ref{inequality2}) is satisfied, periodic distributions instead of the uniform profiles are found to be stable, and thus the striations occur. On the other hand, (\ref{inequality1}) is required for the discharge to be stable: when it is not satisfied, either because the input power is too small (small $r_3$ in (\ref{gainE})) or there is a large recombination loss rate (large $r_2$ in (\ref{lossn})), the discharge is difficult to ignite and is not stable. 
\begin{figure}
\includegraphics[scale=1]{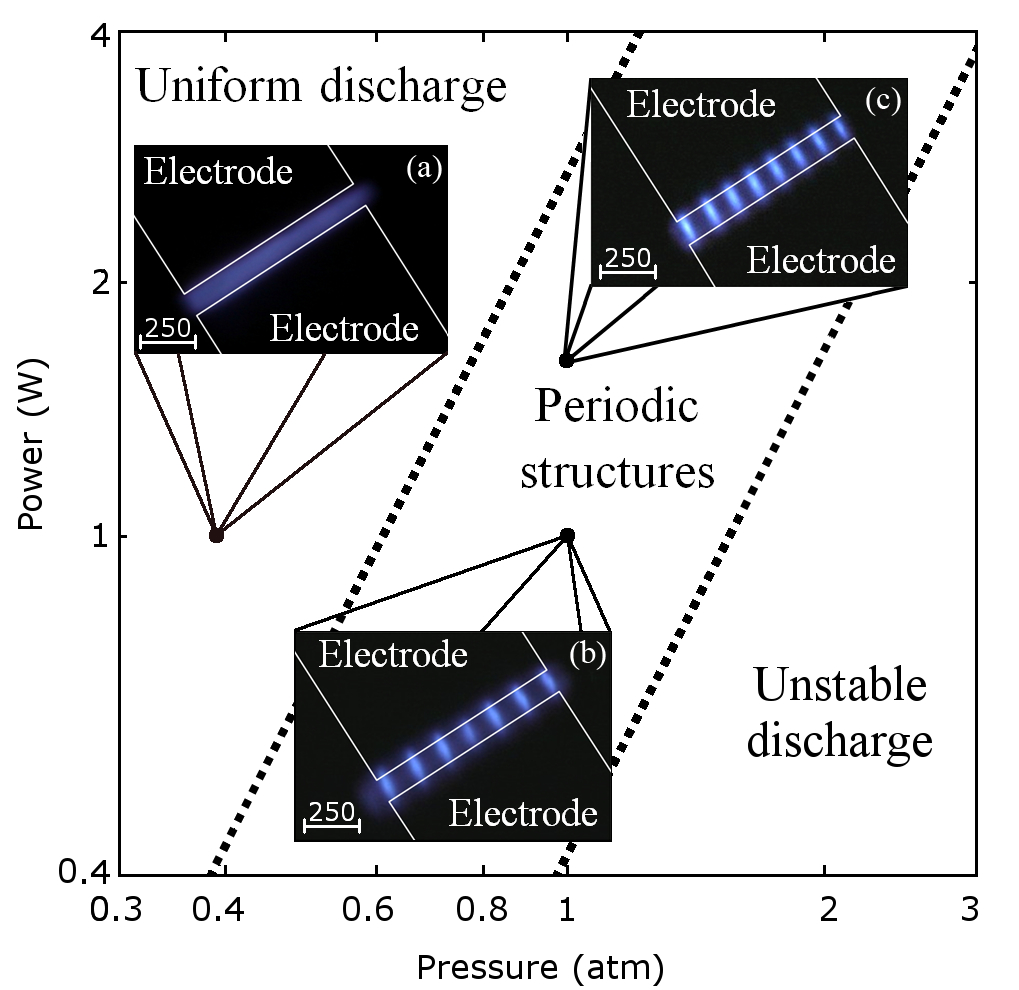}
\includegraphics[scale=0.35]{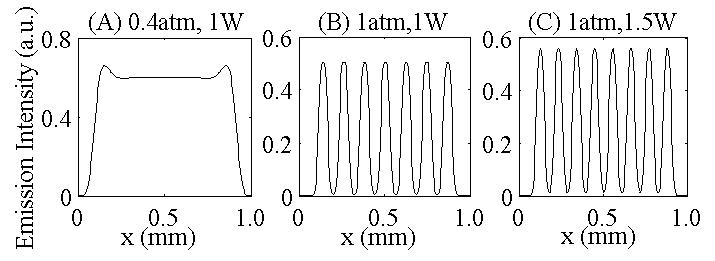}
\caption{\label{fig3}Phase diagram: discharge structure transition as predicted by the proposed model. Images (a)-(c) show experimentally observed distributions of visible light emitted in argon microdischarges (scale bar =250 um) sustained between coplanar electrodes at ~825 MHz at three pressure/power combinations. (A)-(C) are the theoretical results of the emission density distributions with the same pressure/power combination as (a)-(c) respectively.}
 \end{figure}

Based on the physical mechanisms governing the gain and loss terms (\ref{gainn}), (\ref{lossn}), (\ref{gainE}) and (\ref{lossE}), we can assume the parameters in the model scale with pressure $p$ and input power $P$ as follows:
\begin{eqnarray}
&r_1\sim p^1P^0, r_2\sim p^2P^0, r_3\sim p^0P^1, r_4\sim p^1P^0,&\nonumber\\ 
&D_e\sim p^{-1}P^0, D_a\sim p^{-1}P^0, &\label{relation}
\end{eqnarray}
Again, with the parameters of the atmospheric pressure argon discharge \cite{Zhu}, the criteria (\ref{inequality2}) and (\ref{inequality1}) can be presented in a phase diagram, FIG. \ref{fig3}. In the phase on the lower right, where the power is low and the pressure is high, inequality (\ref{inequality1}) is not satisfied and therefore it is not possible to generate stable plasmas under such conditions. The phase in the middle is where the discharge is spatially periodic, and atmospheric pressure gas discharge ($\sim1$ atm, $\sim1$ W) is usually in that phase. If we lower the pressure (to $\sim0.6$ atm) or raise the power  (to $\sim3$ W), we enter the phase on the upper left, where inequality (\ref{inequality2}) does not hold and the discharge is uniform. Despite the simplicity of the proposed Turing model, the phase diagram shown in FIG. \ref{fig3} is in good agreement with experimental observations and therefore suggests that phase transitions between unstable, periodic and uniform discharges can be explained in terms of the activator and inhibitor roles of the electron energy and the electron temperature.

Even though (\ref{gainE}) implies a uniform electric field distribution, the existence of periodic structure solutions obtained from the model is rather insensitive to this assumption. In fact,  numerical simulation results show that  the model can still reproduce  striations, when (\ref{gainE}) is replaced by $G_E\propto n^\gamma$, $\gamma$ varying from 0 to $\sim1.3$. In addition, it can be shown that this approximation does not affect the location of the phase boundary  in FIG. \ref{fig3}, across which the discharge structure changes from uniform to periodic.

In summary, we have shown that striations and discharge structure transitions in atmospheric pressure argon discharges generated  with a microwave source can be predicted by the Turing model. The generality of the physical principles underpinning this model suggests that the proposed model can also be applied to other discharge systems, where the patterns share similar characteristics, such as, the spatial variation  scale lengths decrease as the power increases, and the discharge structure also changed from spatially periodic to uniform when the power is raised or the pressure is decreased \cite{Guikema,Schoenbach}. In fact, in most near-atmospheric pressure gas discharge systems, the production and loss of the excited state atoms are close to local equilibrium due to their small diffusion coefficients. So based on the collisional-radiative model, their densities depend on the local electron density and electron temperature. As a result, the following reaction-diffusion equations can be used to describe the system:
\begin{eqnarray}
&&\frac{\partial n}{\partial t}=D_a\nabla^2n+f(n,T) \label{diffusionn}\\
&&\frac{\partial T}{\partial t}=D_e\nabla^2T +g(n,T)\label{diffusionT}
\end{eqnarray}
Where $f(n,T)$ and $g(n, T)$ include the contribution of the excited state atoms to the production and loss of the electron density and electron energy, respectively. At atmospheric pressure, where the electron temperature is $\sim1$ eV, the stepwise ionization/excitation contributes mostly to the production of electron density/the loss of electron energy, and thus $f$ and $g$ contain nonlinear terms with $\partial f/\partial n>0$ and $\partial g/\partial T<0$. Furthermore, $D_e \gg D_a$ is also satisfied. 
So, the electron density and the electron temperature act as activator and inhibitor in a classical two-reactant Turing model. Therefore similar patterns should be expected in systems where the stepwise ionization/excitation dominate the production of the electron density/loss of the electron energy. Extension of the model is expected to be able to reproduce the selection among different patterns observed in higher dimensional systems, or explain more complicated pattern behaviors, such as the traveling waves. 

The authors would like to thank Y. Lan, Q. Ouyang and D. Economou for helpful discussions. This work is supported by the NSFC Grant No. 11075093 and the National Innovation Research Project for Undergraduates.

\end{document}